 \documentclass[aps,showpacs,superscriptaddress,nofootinbib,amsmath,amssymb,10pt]{revtex4}
 \usepackage{graphicx}
\usepackage{dcolumn}
\usepackage{bm}
\usepackage{lipsum}
\usepackage{adjustbox}
\usepackage{multirow}
\usepackage{amsmath}
\usepackage{amssymb}
\usepackage{gensymb}
\usepackage{textcomp}
\usepackage{enumerate}

\graphicspath{./figs}
\begin{document}
\title{Neutrino Interactions in the SIS and DIS Regions: Current Insights and Future Challenges}
\author{M. Sajjad Athar\footnote{Corresponding author: sajathar@gmail.com\\
\texttt{https://orcid.org/0000-0003-4814-5084}}}
\affiliation{Department of Physics, Aligarh Muslim University, Aligarh - 202002, India}
%

%


\begin{abstract}
In this review, we discuss the current understanding of charged current (anti)neutrino scattering off nucleons and nuclear targets in the few-GeV energy range, a domain of paramount importance for accelerator and atmospheric neutrino experiments. We provide a concise yet comprehensive overview of the experimental and theoretical landscape of neutrino interaction processes across the kinematic region of the shallow and deep inelastic scattering regimes. Moreover, we underscore the pressing unresolved questions and formidable challenges that lie ahead, stressing the urgent need for more refined theoretical models and high-precision measurements to deepen our understanding of neutrino-nucleon and neutrino-nucleus interaction cross sections.
\end{abstract}
\maketitle
\section{Introduction}
Neutrinos, elementary particles of the Standard Model (SM), are massless, electrically neutral, and possess spin  $\frac{1}{2}$. Interacting solely via the weak force, they are the second most abundant particles in the visible universe after photons. Yet, despite their ubiquity, they remain among the least understood entities in modern physics.  Far from being mere spectators in the cosmic play, neutrinos hold the potential to unlock some of the most profound mysteries spanning multiple domains like particle physics, nuclear physics, astrophysics, and cosmology~\cite{SajjadAthar:2021prg}. They challenge our deepest assumptions about symmetry principles and conservation laws, compelling us to rethink the very foundations of the Standard Model.  Over the past several decades, experimental and theoretical efforts have intensified, aiming to unravel their secrets. However, due to their feeble interaction with matter, neutrino experiments rely on moderate to heavy nuclear targets, and future projects envision gigantic nuclear detectors. This necessity places immense emphasis on understanding neutrino interactions with nucleons and nuclear matter, where a better understanding of weak hadronic vertex, nucleon binding effects and nuclear medium modifications play a crucial role.  Studying neutrino interactions is not just an intellectual pursuit, it is essential for advancing our understanding of the universe. The importance of this research can be highlighted through several key points~\cite{SajjadAthar:2022pjt}: 
 \begin{itemize}
  \item [(i)] {\bf{Fundamental to Detection:}} Neutrino interactions form the very foundation of how neutrino detectors operate, determining their ability to observe and measure these elusive particles.
   \item [(ii)] {\bf{Probing the Weak Force:}} They provide a unique testing ground to study the weak nuclear force.
\item [(iii)] {\bf{Deciphering Neutrino Oscillations:}} Precision measurements of these interactions are crucial for uncovering the true nature of neutrino oscillations, mass hierarchy, and potential CP violation in the lepton sector.
\item [(iv)] {\bf{Gateway to New Physics:}} Before we can search for physics beyond the Standard Model, we must first achieve an accurate and comprehensive understanding of neutrino interactions within it.
\end{itemize}
 \begin{figure}
\begin{center}
		\includegraphics[height=6 cm, width=12 cm]{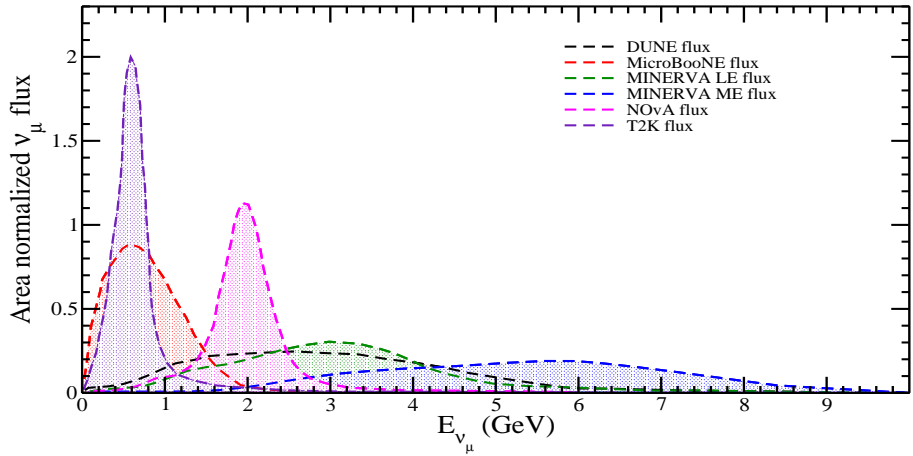}
\end{center}
\caption{The various neutrino fluxes which are being used in the T2K, MINERvA low energy~(LE: $<E_{\nu_\mu}>= 3GeV$), medium energy~(ME: $<E_{\nu_\mu}>= 6GeV$), NOvA, MicroBooNE 
along with the proposed DUNE experiment at the Fermilab are shown to highlight the importance of the understanding of the 
cross section in the few GeV energy region. These neutrino fluxes are normalized to unit area.}
\label{flux}
\end{figure}

\begin{table}
 \begin{tabular}{||c||c||c||}\hline
   \multicolumn{3}{|c|}{Processes}\\\hline
 $CC(W^-)$ &  $CC(W^+)$& $NC(Z^0)$\\\hline
 $\bar\nu_l+p\to l^++X^0$ &  $\nu_l+p\to l^-+X^{++}$ &  $\bar\nu_l(\nu_l)+p\to \bar\nu_l(\nu_l)+X^+$\\\hline
 $\bar\nu_l+n\to l^++X^-$ &  $\nu_l+n\to l^-+X^{+}$ &  $\bar\nu_l(\nu_l)+n\to \bar\nu_l(\nu_l)+X^0$\\\hline\hline
&\multicolumn{2}{|c|}{$X^{++}=p \pi^+,\Sigma^+ K^+$}\\
 &\multicolumn{2}{|c|}{$X^+=p\pi^0, n\pi^+, p\eta,\Lambda K^+,\Sigma^0 K^+, \Sigma^+ K^0$}\\
  $\Delta S=0$  &\multicolumn{2}{|c|}{$X^0=n\pi^0, p\pi^-, n\eta,\Lambda K^0,\Sigma^0 K^0, \Sigma^- K^+$}\\
   &\multicolumn{2}{|c|}{$X^-=n\pi^-, \Sigma^- K^0$}\\\cline{1-3}
   &\multicolumn{2}{|c|}{$X^{++}=\Sigma^+ \pi^+$}\\
 &\multicolumn{2}{|c|}{$X^+=\Lambda^0 \pi^+, \Sigma^+\pi^0, \Sigma^+ \eta, p \bar K^0,\Sigma^0 \pi^+, \Xi^0 K^+$}\\
  $\Delta S=-1$  &\multicolumn{2}{|c|}{$X^0=\Lambda^0 \pi^0, \Sigma^0\pi^0,\Sigma^+ \pi^-,\Sigma^-\pi^+, pK^-, n \bar K^0,\Sigma^0 \eta,\Lambda^0\eta,\Xi^0 K^0,\Xi^- K^+$}\\
   &\multicolumn{2}{|c|}{$X^-=\Lambda^0 \pi^-,\Sigma^0\pi^-,\Sigma^-\pi^0, n\bar K^-,\Sigma^- \eta, \Xi^- K^0$}\\\cline{1-3}
      &\multicolumn{2}{|c|}{$X^{++}=p K^+$}\\
 &\multicolumn{2}{|c|}{$X^+=p K^0,n K^+$}\\
  $\Delta S=+1$  &\multicolumn{2}{|c|}{$X^0=n K^0$}\\\cline{1-3}
 \end{tabular}
     \caption{Charged(CC) and Neutral(NC) current neutrino and antineutrino induced inelastic processes mediated by $W^\pm$ and $Z^0$ bosons rexpectively.}
 \label{sec2:Table1}
\end{table}

  \begin{figure}
 \begin{center}
    \includegraphics[height=7cm,width=12cm]{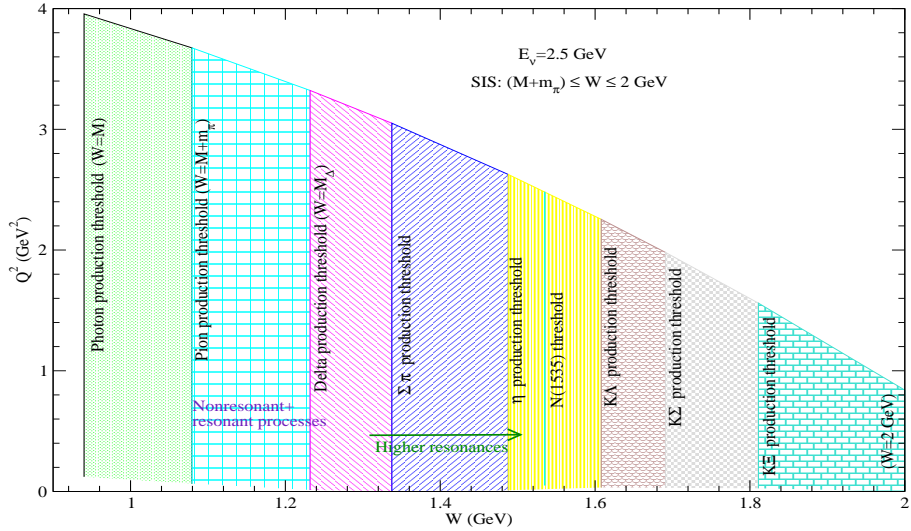}
       \end{center}
       \caption{$Q^2-W$ plane depicting neutrino-nucleon scattering at $E_\nu=2.5$ GeV laboratory neutrino energy.}
       \label{fig1}
 \end{figure}
We shall discuss here interaction of neutrinos in the few GeV energy region which are relevant for the ongoing atmospheric and accelerator neutrino experiments, for example, NOvA, T2K, MINERvA, SuperK, IceCube, etc. or the future experiments such as
DUNE, T2HyperK and nuSTORM. For demonstration, in Fig.~\ref{flux}, area normalized flux for muon type neutrinos determined from some of 
the new generation experiments is shown. In this few GeV energy region, 
the contribution to the cross section comes from different reaction channels such as quasielastic, inelastic (pion production, kaon production, associated particle production, hyperon production, multi-pion production, etc.) and deep inelastic scattering(DIS) processes which have different energy dependence~\cite{SajjadAthar:2024etq}. 
When an (anti)neutrino of four momentum $k$ interacts with a nucleon target of four momentum  $p$ , the center-of-mass energy $W$ is 
given by  $W=\sqrt{(p+q)^2}$, where  $q (=k - k^\prime$, where $k^\prime$ is the outgoing lepton momentum), represents the four-momentum transfer to the hadronic system. The 
interaction corresponds to the quasielastic scattering when  $W=M_N$, $M_N$ is the mass of nucleon. As $W$ increases, additional processes, 
such as single and multiple pion, kaon and eta meson production, begin to contribute. Some of the possible inelastic reactions are given 
in Table~\ref{sec2:Table1}~\cite{SajjadAthar:2022pjt}. Eventually,
the reaction transitions into the deep inelastic scattering regime, where hadronic jets are produced in
the final state.  

To achieve the precise neutrino-nucleon cross sections in the few GeV energy region, it is important to understand the transition region lying between the pion production processes due to the direct interactions as well as 
the resonance excitations 
 and the deep inelastic scattering processes~\cite{NuSTEC:2017hzk}. These extreme regions use different degrees of freedom to describe the neutrino-nucleus interactions.  Among the inelastic scattering processes, the single pion production is predominantly driven by the resonant $\Delta(1232)$ excitation and there is also some contribution from the non-resonant
 and higher-resonant terms. For eta production, the $S_{11}(1535)$ resonance dominates, although 
 additional contributions from non-resonant and higher-resonant terms are also present. With the increase in $W$ and $Q^2(= -q^2)$, 
 contributions from higher resonances become increasingly significant. Some of the higher resonances lying in the second and third resonance regions which contribute to the different meson-baryon production 
 channels are summarized in Ref.~\cite{SajjadAthar:2022pjt}. 
 The need is to understand the vector 
 and axial vector current structure functions in the resonance region, and the transition of the resonance region to the DIS region.

\begin{figure}
 \begin{center}
     \includegraphics[height=7.0 cm,width=12 cm]{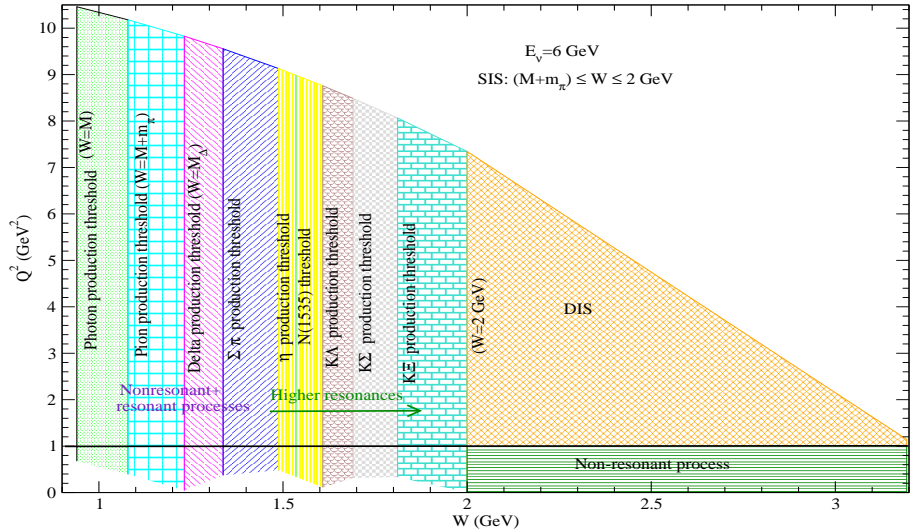} 
       \end{center}
              \caption{$Q^2-W$ plane depicting neutrino-nucleon scattering at $E_\nu=6$ GeV laboratory neutrino energy.}
                 \label{fig2}
 \end{figure}
Figs.~\ref{fig1} and \ref{fig2} illustrate the dependence of $W$ and
$Q^2$ for an (anti)neutrino beam with an energy of 2.5 GeV and 6 GeV, respectively, where the production threshold of different mesons may be observed.
 It is apparent that the resonant and non-resonant pion productions with $W < 2$ GeV overlap and cannot be distinguished. 
 As one moves away from the
 higher $W$ region, where DIS (that deals with the quarks and gluons) is the dominant process to
the region of resonant and non-resonant processes (having hadrons as a degree of freedom), $Q^2\ge 1$ GeV$^2$
 is generally chosen as the lower limit required to be interacting with the hadron's constituents. In the literature, a kinematic constraint
 of $W\ge 2$ GeV is also applied to distinguish the contributions from
 the resonance region and DIS. 
  \begin{figure}
\begin{center}
	\includegraphics[height=7.5 cm, width=12 cm]{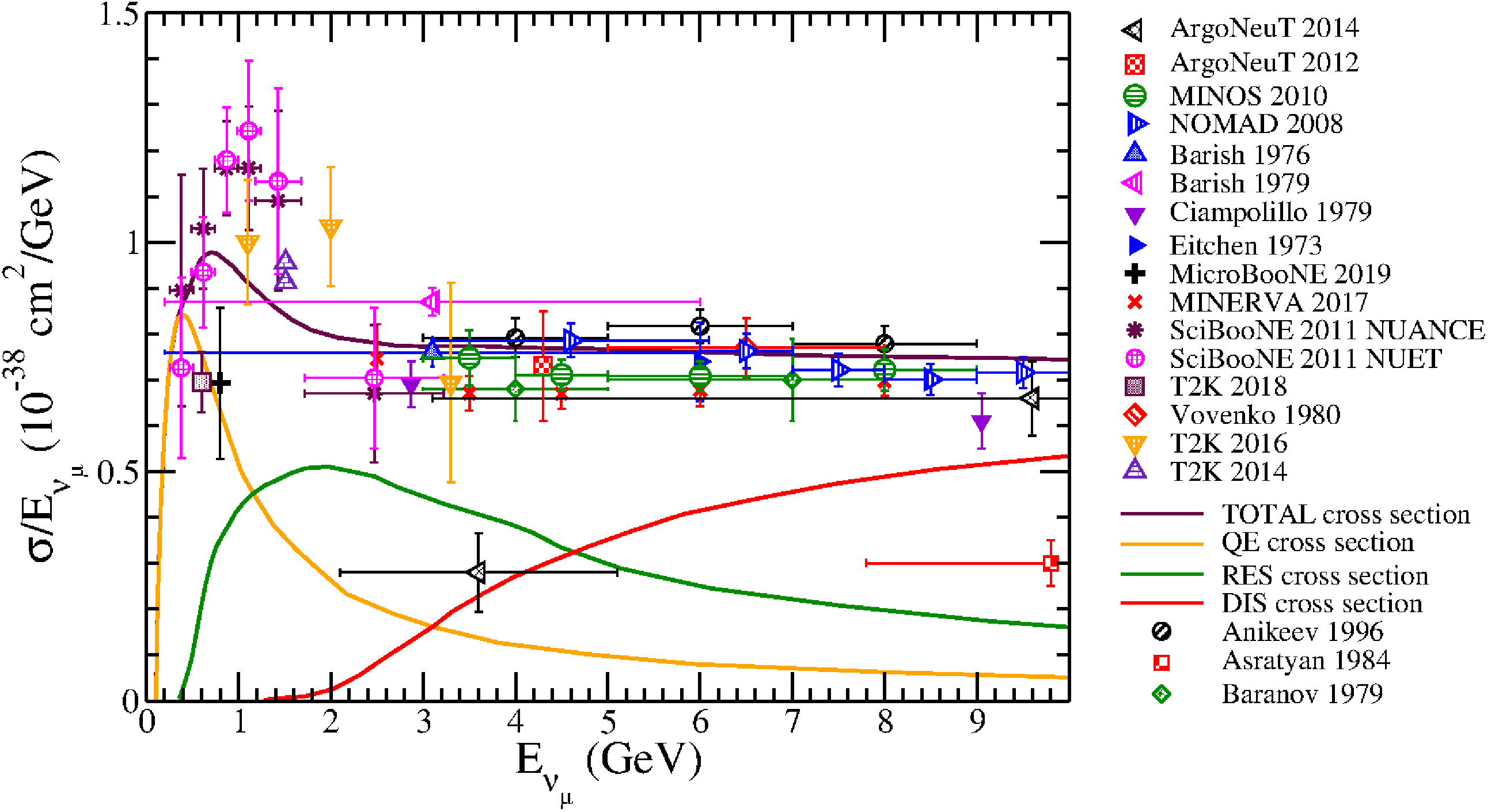}	
\end{center}
\caption{$\frac{\sigma}{E_{\nu_\mu}}$ vs $E_{\nu_\mu}$ for an isoscalar target. The data are the experimental 
points for the inclusive cross section($\sigma$) in various nuclear targets. The theoretical result for $\sigma$(solid 
line) has the contribution from quasielastic scattering~(orange solid line), resonance production~(green solid line), and deep 
inelastic scattering~(red solid line) provided by the NUANCE generator~\cite{Casper:2002sd}.}
\label{xsec_lipari}
\end{figure}
The intermediate region between the meson production threshold and the DIS domain is commonly
referred to in the literature as the shallow inelastic scattering (SIS) region. However, defining the precise 
kinematic boundaries between SIS and DIS is not straightforward, as different experiments employ varying cuts on
$W$. Typically, the SIS region is considered to span from  $W= M_N + M_\pi$ to  $W=2$ GeV. This region represents 
a compelling but not yet fully understood transition from interactions governed by hadronic degrees of freedom to 
those dominated by the quarks and gluons, as described by perturbative QCD~\cite{SajjadAthar:2020nvy}.

Figure~\ref{xsec_lipari} presents the experimental results for inclusive neutrino-nucleus scattering reported by several experiments along with their associated error bars~\cite{Casper:2002sd}. Additionally, the theoretical predictions by Lipari et al.~\cite{Lipari:2002at} for quasielastic, resonance production, deep inelastic, and total scattering cross sections are also shown. In light of the large error bars, precision studies require experiments to reduce systematic errors to the level of a few percent. 
The present experiments face two major sources of systematic errors, both of which significantly impact the accuracy of neutrino interaction studies. The first arises from uncertainties in reconstructing the incident neutrino energy, primarily due to the energy-dependent flux variations of the neutrino beam. The second stems from the challenges in extracting (anti)neutrino-nucleon scattering amplitudes from experimental data on (anti)neutrino-nucleus cross sections.  In the first case, energy reconstruction uncertainties emerge when inferring the incident neutrino energy from the measured energy and angle of the final-state charged lepton, which has scattered quasielastically from an off-shell nucleon inside the nucleus. The nucleon, in motion with a Fermi momentum, introduces an inherent spread in the reconstructed energy. It has been estimated that the lack in the understanding of neutrino-nucleon and neutrino-nucleus
cross sections in the energy range of hundreds-MeV to few-GeV add 25-30\% uncertainty to the systematics.  For example, it has been estimated that for a 1 GeV (anti)neutrino scattering off a nucleon in a  $^{12}C$ nucleus, this uncertainty can be as large as 100 MeV. Furthermore, when considering additional reaction channels beyond pure quasielastic scattering, an energy shift of approximately 100-200 MeV may occur around $E_\nu \sim 1$ GeV, further complicating precise energy determination. The second source of uncertainty arises from nuclear medium effects (NME) on cross sections when (anti)neutrinos interact with bound nucleons. These nucleons, moving within a nuclear potential and subject to Fermi motion, also interact strongly with virtual mesons and neighboring nucleons. This leads to significant contributions from meson exchange currents (MEC) and multinucleon correlations, further distorting the extracted scattering amplitudes~\cite{Fatima:2024hlu}. 
%

 Therefore, it is important to study the kinematic region corresponding to the shallow
inelastic scattering (SIS), where the contribution comes either from the inelastic nucleon resonance region or from the deep inelastic region, in 
order to avoid the double counting of events, which may provide information about the hadron as well as parton degrees of freedom and 
it may also be helpful to define the more accurate kinematic boundaries to demarcate the tail of the nucleon resonance region
and the onset of the DIS region which are not yet uniquely defined in the literature.
SIS region can be understood through the phenomenon 
of quark-hadron duality that bridges the perturbative and nonperturbative QCD regions~\cite{Alvarez-Ruso:2020ezu}. 

In the electromagnetic sector, Bloom and Gilman~\cite{Bloom:1970xb} conducted a detailed analysis of inclusive $e^--p$ scattering data from SLAC and made a remarkable observation that provided a connection between the low-energy and high-energy descriptions of electron-proton scattering. This profound insight, now widely known as Quark-Hadron Duality, establishes a deep connection between resonance excitation at low momentum transfers and DIS at high momentum transfers. Quark-hadron duality asserts that the structure functions governing the interaction at low $Q^2$, when suitably averaged over an appropriate energy interval, are identical to the structure functions measured at high $Q^2$ in the DIS regime. This implies that the underlying physics governing deep inelastic scattering- where partonic degrees of freedom(perturbative QCD)  dominate- is inherently connected to the physics of resonance excitations, where the hadronic picture is more appropriate. Later it was found that duality holds 
good not only over the entire resonance region ($M_N < W < 2$ GeV) known by the name global duality but also for the three prominent resonance regions namely $W=1.2$,
1.5 and 1.7 GeV, a phenomenon known as local duality. This implies that the resonances are not a separate entity but are an intrinsic 
part of the scaling behaviour, thus providing a connection between the behaviour of resonances and scaling to have a common origin in terms of point-like
substructure.  
In the weak sector, there is need of such studies where the interplay between resonant, non-resonant and DIS regions be better understood. It is experimentally challenging to disentangle resonance from non-resonant meson production, therefore, it
is essential to study the transition region where $Q^2$ and $W$ of resonant and non-resonant meson production gradually leads to the DIS region. 

The complexity increases when reactions occur within a nuclear medium. Here, nuclear effects such as Fermi motion,
binding energy, multi-nucleon correlations, and final-state interactions come into play. These medium effects vary significantly for
different processes in the few-GeV energy range and efforts are required to understand them. Therefore, in conclusion, while neutrino physics 
has made significant strides, the number of unresolved questions far exceeds the answers we have found. Addressing these challenges requires a deep and precise understanding of neutrino 
interactions with matter$-$ an essential hurdle that must be overcome before we can truly enter the era of precision neutrino physics.
\section*{Acknowledgement}
M. S. A. is thankful to the Department of Science and Technology (DST), Government of India for providing 
financial assistance under Grant No. SR/MF/PS-01/2016-AMU/G.

\end{document}